\documentclass[useAMS,usenatbib,manuscript]{emulateapj}
\usepackage{epsfig}

\shorttitle{The HI Content of Local Late-Type Galaxies}
\shortauthors{Evoli et al.}

\begin{document}

\title{The H\,I Content of Local Late-Type Galaxies}

\author{C. Evoli\altaffilmark{1}, P. Salucci\altaffilmark{1}, A. Lapi\altaffilmark{2,1} and L. Danese\altaffilmark{1}}
\email{carmelo.evoli@sissa.it}

\altaffiltext{1}{SISSA - Via Bonomea 265 - 34136, Trieste - Italy.}
\altaffiltext{2}{Dip. Fisica, Univ. `Tor Vergata', Via Ricerca Scientifica 1, 00133 Rome, Italy.}

\begin{abstract}
We present a solid relationship between the neutral hydrogen (H I) disk mass and the stellar disk mass of late-type galaxies in the local universe. This relationship is derived by comparing the stellar disk mass function from the Sloan Digital Sky Survey and the H I mass function from the H I Parkes All Sky Survey (HIPASS). We find that the H I mass in late-type galaxies tightly correlates with the stellar mass over three orders of magnitude in stellar disk mass. We cross-check our result with that obtained from a sample of HIPASS objects for which the stellar mass has been obtained by inner kinematics. In addition, we derive the H I versus halo mass relationship and the dependence of all the baryonic components in spirals on the host halo mass. These relationships bear the imprint of the processes ruling galaxy formation, and highlight the inefficiency of galaxies both in forming stars and in retaining their pristine H I gas.
\end{abstract}


\section{Introduction}\label{intro}

During the last years ground- and space-based surveys allowed to probe the physical properties of many millions galaxies both in the local and in the high-redshift Universe. These analysis have been mainly focused on investigating the \emph{stellar} component of galaxies, and have provided us with a much clearer view of when and where star formation occurred along the cosmic time.

However, there is another baryonic component, namely the neutral atomic hydrogen HI, that should be accurately monitored to understand the process of
galaxy formation; in fact, such a component just constitutes the raw material which stars are made of. According to the standard picture, protogalactic halos initially had all the
same cosmological amount of HI gas, around $1/6$ of the host halo mass (e.g., \citealp{komatsu11}), in the form of a warm atmosphere. Then a fraction of
such warm baryons is expected to cool and condense in a cold gaseous disk-like component, whereby stars are formed. In turn, this cold, star-forming gas can
be depleted by the energy feedback from type II supernova explosions and stellar winds, in an amount modulated by the ratio between the total energy
injected and the depth of the potential well of the host halo; noticeably, the former is related to the overall mass of formed stars and hence to the galaxy
luminosity, while the latter crucially depends on the mass of the host halo \citep[e.g.][]{white78,fall80}. At lower halo masses a large   HI depletion is due to   the photo-heating by intergalactic UV radiation field~\citep[e.g.][]{hoeft06,ricotti09}.

Therefore, the observational information on the HI mass content of galaxies provide crucial constraints on galaxy formation theories; a successful scenario must  be able to reproduce not only the observed stellar mass function and luminosity function, but also the HI mass function \citep{mo05} and the relationships between the HI and the stellar/halo mass.

Only in recent years, thanks to the completion of relatively wide blind $21$-cm surveys, a wealth of observations on HI gas has become available. In detail, \citet{zwaan05} used the catalog of $4315$ extragalactic HI $21$-cm emission line detections from the HI Parkes All Sky Survey (HIPASS) to obtain an accurate measurement of the galaxy HI mass function (HIMF) down to an HI mass of $10^{7.2}\,M_\odot$.

In this work we aim at investigating the relationships between the HI mass and two relevant structural properties of late-type galaxies: the stellar disk
and the halo masses. To reach this we exploit: i) a theoretical approach that boils down to matching the cumulative HIMF mass function and the galactic stellar
(or halo) mass function; ii) an observational approach that relies on a sample of objects for which both the HI and stellar disk masses have been
{\it directly} measured. We show that the two approaches agree in indicating a strong correlation between the gaseous disk and stellar disk (or halo) mass.

The existence, in late-type spirals, of a relationship between the HI disk mass and the galaxy luminosity is well known \citep[e.g.][]{roberts75,roberts94,gavazzi96,mcgaugh97,disney08}, so as that between the former and  the spectro-photometrically derived mass of the stellar disk \citep{kannappan04,gavazzi08,catinella10}. Recently, \cite{cortese11} showed that HI-to-stellar mass ratio anti correlates with stellar mass over $\sim$2 order of magnitudes in stellar mass and investigated the effect of the environment on this relation.
 
Previous results, based on the spectrophotometric estimate of the stellar disk masses, have estabilished the {\it existence} of the relations subject of the present investigation, but  in a biased way. In fact, especially for spirals, the luminosity is a poor indicator of the stellar disk mass, and, in any case, it is uncertain by a factor two \citep{salucci08}. In addition, it depends on the assumed initial mass function (IMF) and star formation rate, quantities that we would like to study helped by  the results of this paper and not to assume a priori to get the results of this paper. Finally, the above relationships are biased by the fact that spirals with the same stellar disk mass, but overabundant or deficient in HI content,  seem to have different stellar mass-to-light ratios (and then luminosities) than galaxies ``normal'' in HI content. 
 
In this work, we aim to estimate the mass of a stellar disk, associated to a HI disk in two essentially new, accurate, model-independent and statistically relevant ways. These estimates are expected to yield trustable relationships or trends, free from biases that are likely to affect their interpretation in a cosmological context. Notice that \citet{shankar06}  by following \citet{salucci99}, were the first to correlate the kinematical bias-free estimates of stellar disk mass with the corresponding HI masses, however, their work was based on a sample much more limited, in number of objects and magnitude extension, than that we use in this work. 
 
Finally, we apply the cumulative technique to derive the relationship between HI and halo masses. Even if not strongly motivated as in the previous case, we are able to derive a more realistic relationship for these two observables with respect to what existing in literature.
 
Throughout the paper we adopt the standard value $H_0 = 73$ km s$^{-1}$ Mpc$^{-1}$ for the Hubble constant, and quote uncertainties at $1-\sigma$ confidence level.

\section[]{HI vs. stellar mass relationship}\label{method}

\begin{figure}
\centerline{\psfig{figure=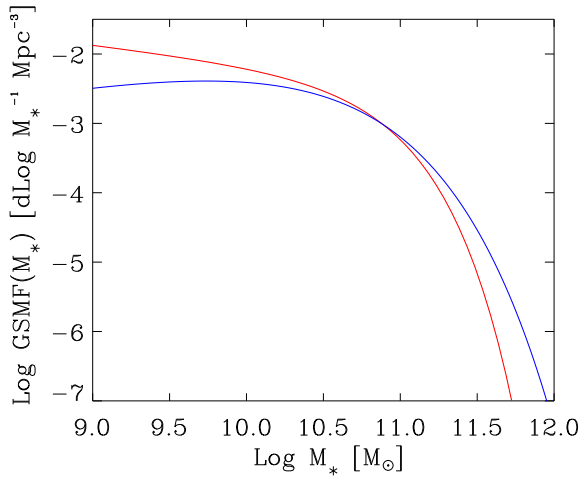,width=\columnwidth,angle=0}}
\caption{The GSMF of late-type galaxies obtained from the LF by \citet{bernardi10} and the M/L ratios by \citet{shankar06} is illustrated as a blue line. The GSMF by \citet{bell03} is also shown  (red line)  for comparison.} \label{GSMF}
\end{figure}

 To investigate the relationship between the stellar and the gas mass component in late-type galaxies, we follow the procedure by \citet{vale04} and developed by \citet{shankar06}. First, supported by the evidence  described in  section .~1 we assume that, in average, the mass of the HI disk is, in statistical sense, an (increasing) monotonic function of the mass of the stellar disk. 

If two galaxy properties $q$ and $p$ obey a one-to-one relationship, we can write:
\begin{equation}
\phi(p)\, \frac{{\rm d} p}{{\rm d} q}\, {\rm d} q = \psi (q)\, {\rm d} q
\end{equation}
where $\psi(q)$ is the number density of galaxies with measured property between $q$ and $q + {\rm d} q$ and $\phi(p)$ is the corresponding number density for the variable $p$. The solution is based on a numerical scheme imposing that the number of galaxies with $q$ above a certain value $\bar{q}$ must be equal to the number of galaxies with $p$ above $\bar{p}$, i.e., 
\begin{equation}\label{maineq}
\int_{\bar{p}}^{\infty}\, \phi(p)\, {\rm d} p = \int_{\bar{q}}^{\infty}\,
\psi(q)\, {\rm d} q ~.
\end{equation}
In the following we take $p$ as the HI mass $M_{\rm HI}$ and $\phi(p)$ as the corresponding HIMF, while $q$ as the stellar mass $M_\star$ and $\psi(q)$ as the corresponding galactic stellar mass function GSMF.
\begin{figure}
\centerline{\psfig{figure=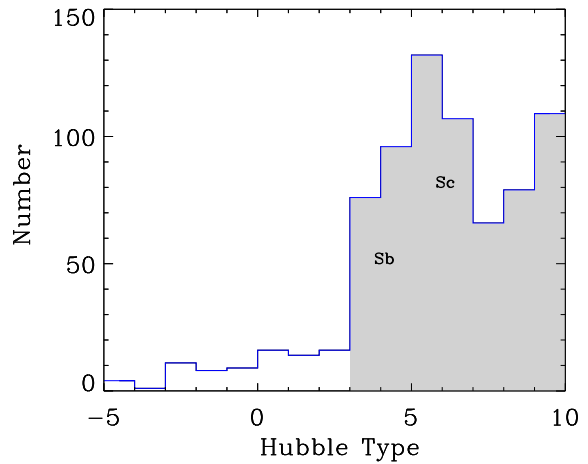,width=\columnwidth,angle=0}}
\caption{The distribution of Hubble-types in the HIPASS galaxies as presented
by \citet{ryan02}.}\label{HT}
\end{figure}
The local HIMF has been measured by \citet{zwaan05} using the galaxy data in the HIPASS catalog \citep{meyer04}; its shape has been fitted, within the range $10^{7.2}\, M_\odot<M_{\rm HI}<10^{11}\,M_\odot$,  with a Schechter function:
\begin{equation}
\phi(M_{\rm HI}) {\rm d}M_{\rm HI} = \phi_{\rm HI} \left( \frac{M_{\rm HI}}{\tilde{M}_{\rm HI}} \right)^{\alpha} \exp \left( -\frac{M_{\rm HI}}{\tilde{M}_{\rm HI}} \right) {\rm d}\left( \frac{M_{\rm HI}}{\tilde{M}_{\rm HI}} \right)
\end{equation}
with power law slope $\alpha = -1.37 \pm 0.03$, characteristic mass $\log (\tilde{M}_{\rm HI}/M_\odot) = 9.8 \pm 0.03 h_{75}^{-2}$ and normalization $\phi_{\rm HI} = (6 \pm 0.8) \times 10^{-3} h_{75}^{3}$ Mpc$^{-3}$ dex$^{-1}$.

Obviously, the two mass functions appearing in Eq.~(\ref{maineq}) must be representative of the same galaxy population. To check this, we plot in Fig.~\ref{HT} the Hubble-type distribution (obtained from the HyperLeda Catalogue, see \citealt{paturel03}) of the $1000$ brightest HIPASS galaxies as reported in \citet{ryan02}. We conclude that the HIMF represents almost entirely disk systems: late-type galaxies account for more than 85\% (Sb-Sc), there is a small contribution from irregular galaxies (smaller than $15\%$), and the contribution from ellipticals is negligible (smaller than $2\%$).

Thus we calculate the GSMF for late-type and Irregular galaxies on the basis of the recent observational results reported in \citet{bernardi10}. Specifically, we use their LF for $C_r < 2.6$ \footnote{$C_{r}$ is the concentration index defined as the ratio of the scale which contains 90\% of the Petrosian light in the $r$ band, to that which contains 50\%.} (M. Bernardi, private communication), which implies a small contamination from early-type galaxies, around $2\%$ from ellipticals and less than $26\%$ from Sa-type objects. From this we build the GSMF by adopting the {\it disk} Mass-to-Light ratio derived from mass modelling of the (Spiral) Universal Rotation Curve, see Eq. 2 in \citet{shankar06}, and we fit it with a modified Schechter function (see \citealt{bernardi10}, Eq.~$9$):
\begin{equation}
\phi(M_{*}) {\rm d}M_{*} = \phi_* \left( \frac{M_{*}}{\tilde{M}_{*}} \right) \frac{{\rm e}^{-(M_{*}/\tilde{M}_{*})^{\beta}}}{\Gamma (\alpha/\beta)} \beta {\rm d}\left(  \frac{M_{*}}{\tilde{M}_{*}} \right)
\end{equation}
with parameters: $\phi_{*} = 1.05 \times 10^{-2}$~Mpc$^{-3}$, $\alpha = 0.385$, $\beta = 0.59$ and $\log M_{*} = 10.05$. The function is plotted in Fig.~\ref{GSMF} alongside, for the sake of comparison, with the GSMF of late-type galaxies obtained by \citet{bell03} from model-dependent spectrophotometric estimates of the disk masses. The method we use suffers for different uncertainties, in particular in inferring stellar masses from kinematical measurements, hence the total uncertainties on our results is of the order of 30\%.
 
\begin{figure}
\centerline{\psfig{figure=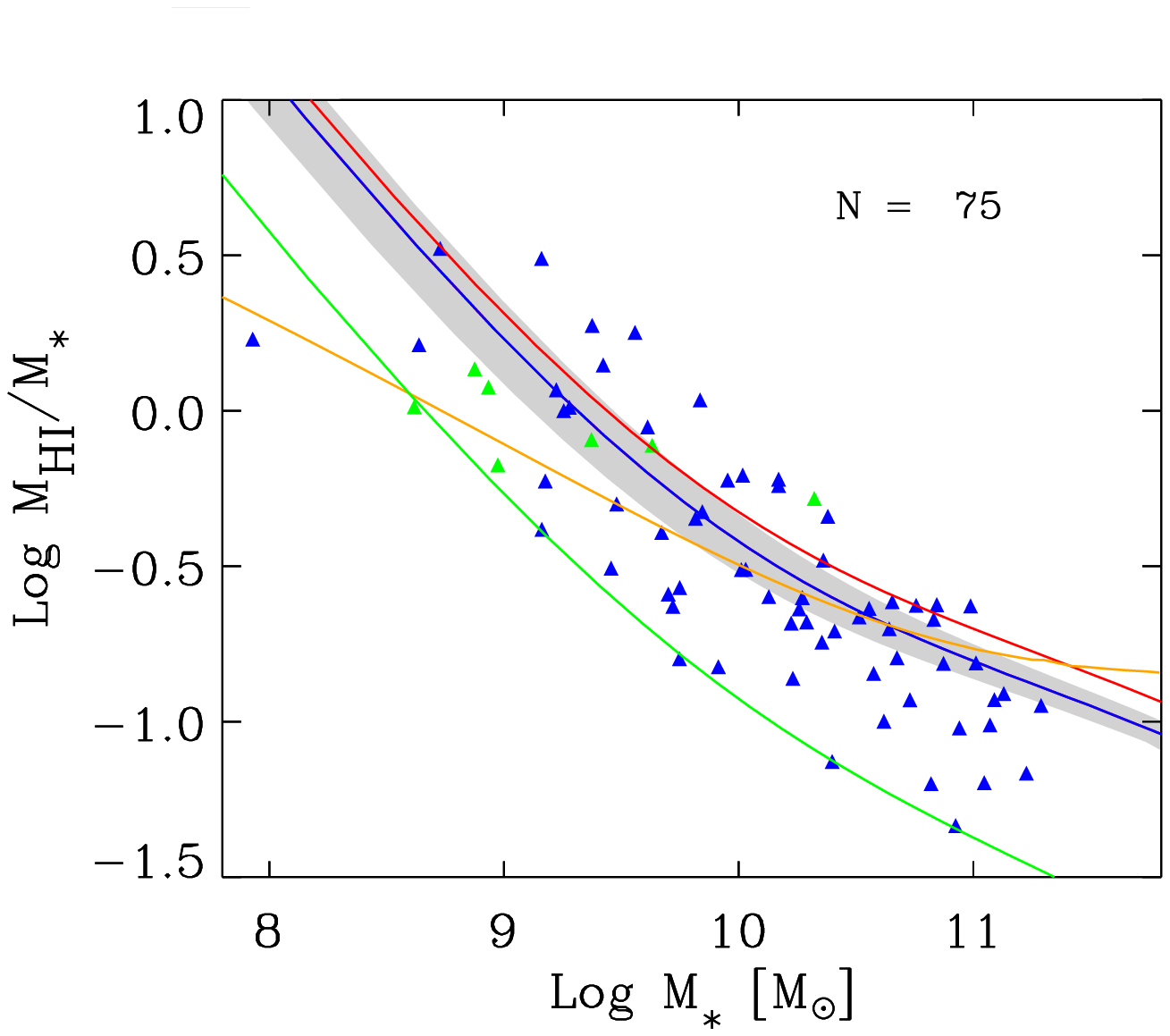,width=\columnwidth,angle=0}}
\caption{Ratio of HI to stellar disk mass as function of the latter, the dashed area represents the uncertainty related to HIMF normalization. Triangles represents individual objects  (blue symbols are for late-types and green for irregulars). Red and orange lines show the effects of changing the HIMF and the GSMF as described in Sec. 2, respectively. The H$_2$ to stellar mass ratio is also reported as a green line.}\label{MHI_Mstar}
\end{figure}

More recently the ALFALFA collaboration have published an HIMF based on 10119 galaxies by probing a bigger volume than HIPASS \citep{martin10}.
The new HIMF differs from the HIPASS one at the high mass end, to the effect of changing the normalization of the HI-to-stellar mass ratio. We show in Fig.~\ref{MHI_Mstar} that the differences in assuming the ALFALFA HIMF are within the error bars associated to the uncertainties in the HIPASS HIMF normalization.

Then,  we solve Eq.~(\ref{maineq}) and derive the relationship between the gas to star fraction and the stellar mass; the result is shown in Fig.~\ref{MHI_Mstar}. The gas fraction and the stellar mass correlates  as a broken power-law over about three order of magnitudes in stellar mass. Within the  mass range $10^{8}<M_{\star}<10^{11}$ the relationship can be well approximated  by :
\begin{equation}\label{MHI:eq}
{M_{\rm HI} \over {3.36\! \times\! 10^9} M_\odot} \!=\! \left(\! \frac{M_\star}{3.3\! \times\! 10^{10} M_\odot} \!\right)^{0.19} \!\left[\!1\!+\!\left(\!\frac{M_\star}{3.3\! \times\! 10^{10} M_\odot} \!\right)^{0.76}\right]
\end{equation}
This relationship, obtained by direct estimate of the stellar disk mass, can be compared with that obtained by means of the (biased) traditional methods. In Fig.~\ref{MHI_Mstar} we compare our result with the HI to stellar mass obtained by using the \citet{bell03} GSMF. The difference between the two is particularly pronounced at small masses, where the spectrophotometric M/L ratios of \citet{bell03} are appreciably larger than the kinematical estimates.

\section{HI content of individual galaxies}

We derive the relationship between the HI mass ($M_{\rm HI}$) and the stellar disk mass ($M_{\rm D}$) with a new model-independent method by looking at {\it individual} late-type galaxies. The disk mass is obtained, within a reasonable uncertainty, by modeling the galaxy rotation curve, whose inner parts are decomposed in halo and disk components. 

Let us first define $R_{\rm opt}\equiv 3.2 R_{D}$, where $R_D$ is the exponential thin disc length-scale. This radius, that encloses about $83\%$ of the total light, can be considered the physical size of the stellar disc. \citet{persic90} devised a reliable method to estimate the disk mass from observational quantities, i.e. from the gravitating mass $M_g$ inside $R_{\rm opt} $ ($M_{g} \approx G^{-1} V_{\rm opt}^2\, R_{\rm opt}$) and the rotation curve logarithmic slope at $R_{\rm opt}$ ($\nabla$): 
\begin{equation}\label{eq:mdisk}
M_{\rm D}= (0.72 - 0.85\, \nabla)~M_g ,
\end{equation}
We then proceed to build a sample containing the $75$ objects in HIPASS that have optical photometry and kinematics of quality sufficient for the above method. The rotation curves are taken from \citet{persic95}, \citet{yegorova07} and \citet{frigerio07}. By means of Eq.~\ref{eq:mdisk} we derive the disk mass with an uncertainty between 10\% -30\% \citep{persic90}.

In Fig.~\ref{MHI_Mstar} we show the relation obtained for individual objects and  that obtained by matching the HIMF to the GSMF. The two are in very good agreement over two order of magnitudes in stellar mass, showing the same power law functional form (with slope respectively of $-0.48$ and $-0.52$) and similar normalization.
The agreement of the individual objects and the statistical relation, obtained from two very different methods, indicates that the first one is little biased by contamination or incompleteness of the HIMF, and that  the second uses a fair sample of individual objects.
A stellar disk mass versus HI disk mass relation emerges as one of the most important empirical relationships concerning spirals.

A further gas component in the local galaxies is the molecular hydrogen (H$_2$) disk. Although we must caveat that its mass does not necessarily monotonically correlate with the stellar disk mass \citep[e.g.][]{casoli96,boselli02,boker03}, we will proceed as above, since more  information on this poorly known component is  certainly needed. 
 
Let us stress that, unlike the HI mass, the H$_2$ disk mass estimate relies on indirect tracers as CO lines, with uncertain conversion factors. We adopt the H$_{2}$MF derived by \citet{obreschkow09} from the local CO luminosity function of the CARO Extragalactic CO Survey, assuming a variable CO to H$_2$ conversion factor fitted to nearby observations. The corresponding mass function (H$_{2}$MF) is well fitted by a Schechter function with powerlaw slope $\alpha = -1.07$, characteristic mass $\log (\bar M_{\rm H_2}/M_\odot) = 9.2$ and normalization $\phi_{\rm H_2} = 8.3\times 10^{-3}$ Mpc$^{-3}$ dex$^{-1}$. The resulting H$_2$ to stellar mass ratio as a function of the stellar mass is shown in Fig.~\ref{MHI_Mstar}: as expected, this component turns out to be subdominant relative to HI over the whole probed mass range and for this reason we do not consider this contribution in the rest of the paper.

\section{HI vs. halo mass relationship}

It cosmologically relevant to derive the relationship between HI mass $M_{\rm HI}$ and halo mass $M_{\rm H}$ in spirals. A preliminary step is to obtain the relationship between the stellar mass $M_\star$ and halo mass $M_{\rm H}$ by the method described in Sect.~2. \citet{shankar06} already obtained this results but it is worth to redo their analysis with updated observational data.
\begin{figure}
\centerline{\psfig{figure=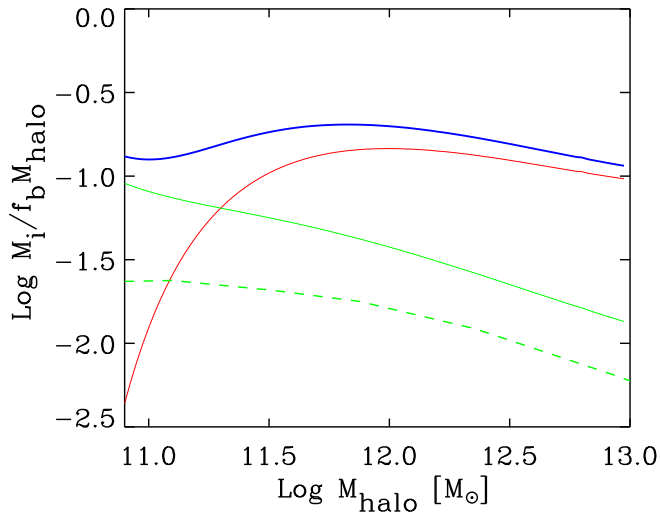,width=\columnwidth,angle=0}}
\caption{The baryonic content relative to the initial baryonic mass
associated to a halo ($f_{b}M_{\rm halo}$). Red line refers to stars, green line to HI 
and blue to the total. Dashed line shows the finding by \citet{marin10}.} \label{bar}
\end{figure}
To this purpose, we need two ingredients. The first is the galactic halo mass function (HF), i.e., the statistics of halos containing one single galaxy; \citet{shankar06} evaluated it from the standard halo mass function by adding the contribution of subhalos, and subtracting the contribution of galaxy systems (see their Eq.~9). 

The second is the GSMF of all the local galaxy population, necessary because the HF does not distinguish between galaxy morphology. We base on the GSMF by \citet{bernardi10} without selection criteria in concentration index; this is fitted in terms of a modified Schechter function (see their Eq.~9) with parameters given in their Table B5.

The relationship derived with these mass functions holds for the overall galaxy population, so to proceed further we must assume that it also approximately holds for each separate Hubble type, in particular, for {\it late-type} objects.  This is justified by the fact that we found that the fractional amount of the HI component with respect to the whole baryonic component vary across Spirals by 3 orders of magnitudes; on the other hand, from X-ray and weak lensing observations, we can infer that galaxies with the same halo mass have approximately the same baryonic mass and that, furthermore, the relation between the galaxy virial mass and the relative baryonic mass is roughly Hubble Type independent (\citealp{fukazawa06}, \citealp{nagino09}, \citealp{donato09}). 

Thus, we combine the HI vs. stellar mass relationship with the stellar vs. halo mass relationship to obtain the HI vs. halo mass relationship. We
show the result in Fig.~\ref{bar}; the relation can be fitted (to better than $5\%$ comparing with the numerical result) within the mass range $10^{11}\, M_\odot <M_{\rm H}<10^{12.5}\,M_\odot$ as:
\begin{equation}
{M_{\rm HI}\over 9 \times 10^9\,M_\odot}= {(M_{\rm H}/9.5\times 10^{11}\,
M_\odot)^{0.33} \over 1+(M_{\rm H}/9.5\times 10^{11}\,M_\odot)^{-0.77}}~.
\end{equation}
We also plot for comparison the HI vs. halo mass relationship recently derived by \cite{marin10} by comparing directly the statistics of HI and halo masses. Their results appreciably differs from ours since the standard halo mass function they adopt includes the contribution of galaxy groups systems so it has  more objects relative to our GHMF; then,  the matching procedure of Sect.~2 leads  to a lower HI mass at a given halo mass.

In Fig.~\ref{bar} we summarize our results by showing the amount of HI and stellar mass (relative to the initial baryonic mass )associated with a halo as a function of its virial  mass. We also plot the overall baryon fraction derived by adding the stellar mass to the total gas mass obtained by multiplying the HI mass for $1.4$ to take into account the contribution of He.

\begin{figure}
\centerline{\psfig{figure=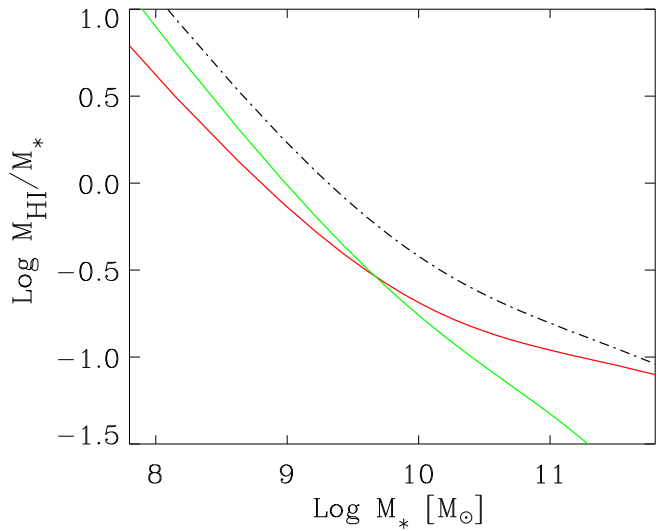,width=\columnwidth,angle=0}}
\caption{In red the HI mass inside the optical radius, in green the HI mass outside, and in black the total.} \label{inandout}
\end{figure}

\section{Discussion and conclusions}

The correlations of the HI mass with stellar and halo masses are extremely relevant in the framework of galaxy formation theories. The standard picture envisages that every galaxy forms with the same initial amount of baryon in the form of HI gas, and what we observe now is the left-over of the processes that took place during galaxy formation.

Fig.~\ref{bar} shows that late-type galaxies are extremely inefficient in retaining their initial baryon content, i.e., most of the initial HI gas has been removed from the host halo. Less than $10\%$ is retained in galaxies with halo masses below $10^{11}\, M_{\odot}$, and this value drops to few $\%$ for halo masses above $10^{12}\, M_\odot$. Such a behavior is likely due to supernova feedback. Thus only a small fraction of the initial baryon content is eventually exploited for the star formation. Note that, in massive halos, stars are the dominant baryonic component while in smaller halos HI gas is.

Let us stress however that the baryon cycle in spirals is very complicated to understand. It may depend, in addition to SN feedback, IGM ionization, gas cooling time, on the interplay between galaxies and their environment, especially for low mass halos. Notice SPH simulations have not yet converged to a definitive result, e.g.~\citet{hoeft06} finds  that halos with $M>10^{10.5}$~M$_{\odot}$ are able to retain all their baryons while \citet{pilkington11} find the galaxy formation process able to remove most of the original baryonic material. All this  means that the processes that are responsible of the evolution of galactic gas about  which this paper  provides valuable information are not fully understood.

One can wonder why this gas, although not being ejected by supernova feedback, has not been used for star formation. To answer the question, we look at where  this residual HI gas is presently residing by highlighting in the previous correlations the contribution from the HI gas located inside or outside the stellar disc radius $R_{\rm opt}$. Therefore,  we model the gas surface density of  late-type galaxies with the functional form observed in most   Spirals \citep[see e.g.][]{bigiel10}
\begin{equation}
\log \Sigma =
\left\{
\begin{array}{ll}
\log \Sigma_0 & \mbox{if $r \leq R_{\rm opt}$}\\
\log \Sigma_0 - 2\, (r-R_{\rm opt})/(R_{\rm f}-R_{\rm opt}) & \mbox{if $r > R_{\rm opt}$}
\end{array}
\right.
\end{equation}
where $R_{\rm f}$ is the radius at which the surface density drops at $1/100$ of the value at $R_{opt}$ that we assume as the size of the HI disk and $\Sigma_0$ is the HI  surface density central value. 

We need now to relate the lenght-scale of the stellar distribution with that of the neutral gas. Notice that our aim is to obtain qualitative results, in this view the assumptions we take are well justified. \citet{broeils97} and \citet{rhee96} published the HI surface density profiles for 60 spirals of known optical radii $R_{\rm opt}$ and blue luminosity $L_B$ (that are given in Tab.~1 of \cite{rhee96}, notice that the quantity in the fourth column is $\simeq R_{\rm opt}/2$). From these measurements they derived: a) the HI half-mass radius $R_{\rm eff}$; b) the total mass $M_{\rm HI}$ (given in columns 5 and 6 of the same Table). From these quantities we obtain a strong $R_{\rm eff}$ vs $R_{\rm opt}$ relationship and, by the definition of $R_{\rm eff}$, the relationship $R_f= F(R_{\rm eff} (R_{\rm opt}), R_{\rm opt})$. Moreover, to transform light in stellar mass, we use, without loss of generality, $M_\star = 10^{11} ( L_B/10^{11} L_\odot )^{1.4} M_\odot$ \citep{salucci07}. Finally by combining and manipulating the above empirical relationships (that also imply to assume Eq.~8) we obtain:
\begin{equation}
{R_{\rm f} / R_{\rm opt}} = 3 - 2/3 \, \log \left( {M_\star} / {10^9 M_\odot} \right) \, .
\end{equation}
The above indicate, not  surpisingly , that small galaxies have a larger HI disk,  in terms of the stellar disk size.
In Fig.~\ref{inandout} we show how the HI   mass   is divided  in those inside and outside $R_{\rm opt}$, the radius inside which the stars reside. The former is the dominant component for massive objects, while the latter gives a dominant contribution in small galaxies.

The overabundance of HI over stellar mass in small objects,  is due to material   located far away the stellar disc and mostly unprocessed. It is worth noticing that in these objects at these radii the  HI  surface density is much lower than the threshold of order $1\, M_{\odot}$ kpc$^{-2}$ needed by the Toomre criterion to form stars. This HI component has not been at disposal for the latter process and it never will. Let us stress  that  the inefficiency of the star formation process in the outer regions of discs is directly probed \citet{bigiel10}, 

To sum up, in this work we have derived robust correlations between the HI and stellar (halo) mass for late-type galaxies in the local Universe. These relationships bear the imprint of the processes ruling galaxy formation (see \citealt{Cook10} for a theoretical approach that consider them), and highlight the inefficiency of galaxies both in forming stars and in retaining their pristine HI gas.

\section*{Acknowledgments}
We acknowledge M. Bernardi for having provided us the LF data for late-type galaxies.
Work partially supported by MIUR, INAF and ASI. A.L. thanks SISSA and INAF-OATS for warm hospitality.

\bibliographystyle{apj}
\bibliography{ms}

\end{document}